      [1999/12/01 v1.4c Il Nuovo Cimento]
\documentclass{cimento}

\usepackage{graphicx}  
\usepackage{amsmath,amssymb}
\title{More on the relation between the two physically inequivalent \\
decompositions of the nucleon spin}
\author{M.~Wakamatsu\from{ins:x}} 
\instlist{\inst{ins:x} Department of Physics, Faculty of Science, 
Osaka University, Osaka, JAPAN}
\begin{document}


\maketitle

\begin{abstract}
The recent controversy on the nucleon spin decomposition
problem is critically overviewed.
We argue that there exist two and only two physically
inequivalent gauge-invariant decompositions of the longitudinal nucleon spin,
contrary to the rapidly spreading view in the QCD spin physics community
that there are infinitely many decompositions of the nucleon spin.
\end{abstract}

\section{Introduction}
Is a gauge-invariant complete decomposition of the nucleon spin possible ? 
It is a fundamentally important question of QCD as a color gauge theory.
The reason is that the gauge-invariance is a {\it necessary condition} of
{\it observability}. Unfortunately, this is quite a delicate problem,
which is still under intense debate.
We feel that the recent INT workshop on gOrbital Angular Momentum
in QCD'' increased controversy rather than settled it.
We therefore believe it an urgent task to correct widespread misunderstanding
on the meaning of {\it true} gauge-invariance in the nucleon spin decomposition
problem.  


\section{Decomposition of gauge field into physical and pure-gauge components}

In a series of papers \cite{ref:Waka10}\nocite{ref:Waka11A}
\nocite{ref:Waka11B}-\cite{ref:Waka12},
we have established that there are two physically
inequivalent gauge-equivalent decompositions of the nucleon spin, which we
call the decomposition (I) and (II). The decomposition (I) and (II) are
respectively characterized by two different orbital angular momenta (OAMs)
for both of quarks and gluons, i.e. the ``dynamical" OAMs and the generalized
``canonical" OAMs. The basic assumption for deriving these two
gauge-invariant decompositions of the nucleon spin is that
the total gluon field can be decomposed into the two parts as 
\begin{eqnarray}
 A^\mu (x) \ = \ A^\mu_{phys} (x) \ + \ A^\mu_{pure} (x), \label{eq:cvdecomp}
\end{eqnarray}
satisfying the following conditions, i.e. the pure-gauge condition
for the pure-gauge component of $A^\mu$,
\begin{equation}
 F^{\mu \nu}_{pure} \ \equiv \ \partial^\mu \,A^\nu_{pure} \ - \ 
 \partial^\nu \,A^\mu_{pure} \ - \ i \,g \,
 [\, A^\mu_{pure}, A^\nu_{pure} \,] \ = \ 0, \label{eq:cond1}
\end{equation}
and the transformation properties for the physical and
pure-gauge components of the gluon field $A^\mu$ given by 
\begin{eqnarray}
 A^\mu_{phys}(x) &\rightarrow&
 U(x) \,A^\mu_{phys} (x) \,U^{-1}(x), \label{eq:cond2} \\
 A^\mu_{pure}(x) &\rightarrow&
 U(x) \,\left(\,A^\mu_{pure}(x) \ + \ \frac{i}{g}
 \,\, \partial^\mu \,\right) \,U^{-1}(x) , \label{eq:cond3}
\end{eqnarray}
under general gauge transformation of QCD.
A question is whether the the conditions (\ref{eq:cond1}),(\ref{eq:cond2})
and (\ref{eq:cond3}) are enough to
uniquely fix the decomposition (\ref{eq:cvdecomp}).
Naturally, the answer is No !
Note however that the decomposition (\ref{eq:cvdecomp}) is proposed as a
covariant generalization of Chen et al.'s decomposition given in a
noncovariant form \cite{ref:Chen08},\cite{ref:Chen09} :
\begin{equation}
 \mbox{\boldmath $A$} (x) \ = \ \mbox{\boldmath $A$}_{phys} (x) \ + \ 
 \mbox{\boldmath $A$}_{pure} (x) ,  \label{eq:ncvdecomp}
\end{equation}
One must know the fact that,
at least in the QED case, this decomposition is nothing
new. It just corresponds to the standardly-known transverse-longitudinal
decomposition of the 3-vector potential of the photon field,
\begin{equation}
 \mbox{\boldmath $A$} (x) \ = \ \mbox{\boldmath $A$}_\perp (x) \ + \ 
 \mbox{\boldmath $A$}_\parallel (x), \label{eq:trans-long-decomp}
\end{equation} 
satisfying the conditions : 
\begin{equation}
 \nabla \cdot \mbox{\boldmath $A$}_\perp \ = \ 0, \ \ \ \ 
 \nabla \times \mbox{\boldmath $A$}_\parallel \ = \ 0.
\end{equation}
It is a well-established fact that this decomposition is {\it unique}
once the Lorentz frame is fixed.
A crucially important ingredient here is the transversality condition
$\nabla \cdot \mbox{\boldmath $A$}_\perp = 0$ for the transverse component
$\mbox{\boldmath $A$}_\perp$. Naturally, an analogous condition
is necessary to uniquely fix the physical component of $A^\mu_{phys}$ in
the decomposition (\ref{eq:cvdecomp}) given in the (seemingly) covariant form.
This fundamental fact of gauge theory is not properly understood in the
community, and conflicting views have rapidly spread around.
On the one hand, Lorc\'{e} claims that the decomposition (\ref{eq:cvdecomp})
is not unique because of the presence of the hidden Sl\"{u}ckelberg-like symmetry,
which alters both of $A^\mu_{phys}$ and $A^\mu_{pure}$ while keeping their sum
intact \cite{ref:Lorce12}.
This misapprehension comes from the oversight of the importance of the
transversality condition that should be imposed on the physical component.
We shall demonstrate this fact in the next section through a simple example
from electrodynamics. 
   
On the other, another argument against the uniqueness of the
decomposition (\ref{eq:cvdecomp}) is advocated by Ji et al. \cite{ref:Ji12}.
According to them, the Chen
decomposition is a gauge-invariant extension (GIE) of the Jaffe-Manohar
decomposition based on the Coulomb gauge, while the Bashinsky-Jaffe
decomposition is a GIE of the Jaffe-Manohar decomposition based on the
light-cone gauge.
They claim that, since the way of GIE is not unique, there is no need that
these two decompositions give the same physical predictions.
This made Ji reopen his longstanding claim that the gluon spin
$\Delta G$ in the nucleon is not a gauge-invariant quantity in a {\it true} or 
{\it traditional} sense, although it is a measurable quantity in
polarized deep-inelastic scatterings.
One should recognize a self-contradiction inherent in this claim.
In fact, first remember the fundamental proposition of physics : 
``Observables must be gauge-invariant."
The contraposition of this proposition (it is always correct if the original
proposition is correct) is gGauge-variant quantities cannot be observables".
This dictates that, if $\Delta G$ is claimed to be observable, it must
be gauge-invariant also in a traditional sense.

\section{The Chen decomposition is not a GIE a la St\"{u}ckelberg}

In this section, we clarify the following two facts in easier QED case
given by the following Hamiltonian, 
\begin{eqnarray}
 H &=& \sum_i \,\frac{1}{2} \,m_i \,\dot{\mbox{\boldmath $r$}}_i^2 \ + \ 
 \frac{1}{2} \,\int \,d^3 r \,
 (\,\mbox{\boldmath $E$}^2 + \mbox{\boldmath $B$}^2 \,),
\end{eqnarray}
which describes an interacting system of charged particles and photons.
First, Chen et al's decomposition is not a GIE {\it a la} Stueckelberg.
Second, there are two and only two physically inequivalent
decompositions of total angular momentum of charged particle and photon
system. Let us start with the expression for the total angular momentum
of this system.
\begin{eqnarray}
 \mbox{\boldmath $J$}
 &=& \sum_i \,\mbox{\boldmath $r$}_i \times m_i \,
 \dot{\mbox{\boldmath $r$}}_i
 \, + \, \int \,d^3 r \,\,\mbox{\boldmath $r$} \times 
 \left(\,\mbox{\boldmath $E$} \times \mbox{\boldmath $B$}\,\right) .
\end{eqnarray}
Here, the 1st term represents the OAM carried by the charged particles,
while the 2nd term does the total angular momentum of the photon.
There is no doubts that the two terms on the r.h.s are both gauge-invariant. 
As already noticed, the vector potential $\mbox{\boldmath $A$}$
of the photon field can be decomposed into longitudinal and transverse
components as (\ref{eq:trans-long-decomp}).
We emphasize again that this
longitudinal-transverse decomposition is {\it unique}, once the Lorentz frame of
reference is fixed. Under a general gauge-transformation given by the
following equations,
\begin{eqnarray}
 A^0 &\rightarrow& A^{\prime 0} \ = \ A^0 \ - \ 
 (\partial \,/\,\partial t) \,\Lambda(x), \hspace{6mm}
 \mbox{\boldmath $A$} \ \rightarrow \ \mbox{\boldmath $A$}^\prime \ = \ 
 \mbox{\boldmath $A$} \ + \ \nabla \Lambda (x) ,
\end{eqnarray}
the longitudinal and transverse components transform as 
follows, 
\begin{equation}
 \mbox{\boldmath $A$}_\parallel \ \rightarrow \ 
 \mbox{\boldmath $A$}^\prime_\parallel \ = \ 
 \mbox{\boldmath $A$}_\parallel \ + \ \nabla \Lambda (x), \hspace{6mm}
 \mbox{\boldmath $A$}_\perp \ \rightarrow \ 
 \mbox{\boldmath $A$}^\prime_\perp \ = \ 
 \mbox{\boldmath $A$}_\perp. \label{eq:GaugeTr}
\end{equation}
This means that $\mbox{\boldmath $A$}_\parallel$ carries unphysical
gauge degrees of freedom, while $\mbox{\boldmath $A$}_\perp$
is intact under gauge transformations. 
To avoid misunderstanding, we emphasize that the above longitudinal-transverse decomposition should clearly be distinguished from the Coulomb gauge fixing.
The Coulomb gauge fixing is to require 
$\nabla \cdot \mbox{\boldmath $A$} = 0$.
Because $\nabla \cdot \mbox{\boldmath $A$}_\perp = 0$ by definition,
this is equivalent to
requiring that $\nabla \cdot \mbox{\boldmath $A$}_\parallel = 0$.
This is the Coulomb gauge
fixing condition. After this gauge choice, $\mbox{\boldmath $A$}_\parallel$ is
divergence-free as well as irrotational by definition, so that one can set
$\mbox{\boldmath $A$}_\parallel = 0$ without loss
of generality.

Another important remark is as follows.
Naturally, the longitudinal-transverse decomposition of the 3-vector potential
is {\it Lorentz-frame dependent}. (Anyhow, the whole treatment above
 is {\it non-covariant}.)
It is true that the Coulomb gauge condition 
$\nabla \cdot \mbox{\boldmath $A$} = 0$ is not preserved, once we move to different Lorentz frame.
Here, we need another gauge-transformation to get vector potential satisfying the Coulomb gauge condition. 
Nonetheless, the Lorentz-frame dependence of the longitudinal-transverse
decomposition does not make any trouble, because one can start this decomposition
in an arbitrarily chosen Lorentz frame.
After all,  the gauge- and frame-independence of  observables is the core of
Maxwell's electrodynamics as a Lorentz-invariant gauge theory !

Now we come back to our original task. As written very clearly in the
text book of electrodynamics \cite{ref:CohenT89}, the total
angular momentum of the photon can actually be split into three gauge-invariant
pieces as,
\begin{eqnarray}
 \mbox{\boldmath $J$}^\gamma \ = \ \int \,d^3 r \,\,
 \mbox{\boldmath $r$} \times (\mbox{\boldmath $E$} \times \mbox{\boldmath $B$})
 \ = \ \mbox{\boldmath $J$}_{long} \ + \ 
 \mbox{\boldmath $J$}_{trans} ,
\end{eqnarray}
with
\begin{eqnarray}
 \ \ \ \mbox{\boldmath $J$}_{long} &\equiv& \int \,d^3 r \,\,
 \mbox{\boldmath $r$} \times 
 (\mbox{\boldmath $E$}_\parallel \times \mbox{\boldmath $B$}) \ = \  
 \sum_i \,q_i \,\mbox{\boldmath $r$}_i \times 
 \mbox{\boldmath $A$}_\perp (\mbox{\boldmath $r$}_i), \\
 \ \ \ \mbox{\boldmath $J$}_{trans} &\equiv&
 \int \,d^3 r \,\,\mbox{\boldmath $r$} \times 
 (\mbox{\boldmath $E$}_\perp \times \mbox{\boldmath $B$}) \ = \
 \int \,d^3 r \,\,E_\perp^l \,
 (\mbox{\boldmath $r$} \times \nabla) A_\perp^l \ + \ 
 \int \,d^3 r \,\,\mbox{\boldmath $E$}_\perp \times \mbox{\boldmath $A$}_\perp.
\end{eqnarray}
Here, $\mbox{\boldmath $J$}_{long}$ is nothing but the
{\it potential angular momentum} in our terminology \cite{ref:Waka10}.
Each term of the above decomposition is separately gauge-invariant, because
$\mbox{\boldmath $A$}_\perp$ is gauge invariant.

What happens if we combine the potential angular momentum term with the ``mechanical"
angular momentum of charged particles ?  We get
\begin{eqnarray}
 \sum_i \,\mbox{\boldmath $r$}_i \times m_i \,\dot{\mbox{\boldmath $r$}}_i \ + \ 
 \sum_i \,\mbox{\boldmath $r$}_i \times q_i \,
 \mbox{\boldmath $A$}_\perp (\mbox{\boldmath $r$}_i) \ = \ 
 \sum_i \,\mbox{\boldmath $r$}_i \times 
 \left(\,\mbox{\boldmath $p$}_i \ - \ q_i \,
 \mbox{\boldmath $A$}_\parallel (\mbox{\boldmath $r$}_i) \,\right) . \label{eq:sum}
\end{eqnarray}
Here, we have used the usual definition of the canonical momentum.
\begin{eqnarray}
 \ \ \ \mbox{\boldmath $p$}_i \ \equiv \ 
 \partial L \,/\,\partial \,\dot{\mbox{\boldmath $r$}}_i \ = \ 
 m_i \,\dot{\mbox{\boldmath $r$}}_i \ - \ 
 q_i \,\mbox{\boldmath $A$} (\mbox{\boldmath $r$}_i)
 \ = \ m_i \,\dot{\mbox{\boldmath $r$}}_i \ - \ q_i \,
 \left(\, \mbox{\boldmath $A$}_\parallel (\mbox{\boldmath $r$}_i) \ + \ 
 \mbox{\boldmath $A$}_\perp (\mbox{\boldmath $r$}_i) \,\right) .
\end{eqnarray}
Note that, on the l.h.s. of (\ref{eq:sum}), the $\mbox{\boldmath $A$}_\perp$ terms
cancel out and $\mbox{\boldmath $A$}_\parallel$ remains.

This leads to a gauge-invariant decomposition  corresponding to Chen
et al.'s.
\begin{eqnarray}
 \mbox{\boldmath $J$} \ \ = \ \ \mbox{\boldmath $L$}^\prime_p 
 \ + \ \mbox{\boldmath $S$}^\prime_\gamma 
 \ + \ \mbox{\boldmath $L$}_\gamma^\prime ,
\end{eqnarray}
where
\begin{eqnarray}
 \mbox{\boldmath $L$}^\prime_p &=& \sum_i \,\mbox{\boldmath $r$}_i \times
 (\mbox{\boldmath $p$}_i - q_i \,
 \mbox{\boldmath $A$}_\parallel (\mbox{\boldmath $r$}_i))
 \ \Rightarrow \ \sum_i \,\mbox{\boldmath $r$}_i \,\times \,
 (\,1 \,/\,i \,) \,\mbox{\boldmath $D$}_{i,pure}, \\
 \mbox{\boldmath $S$}^\prime_\gamma &=& \int \,d^3 r \,
 \mbox{\boldmath $E$}_\perp \times \mbox{\boldmath $A$}_\perp, \\
 \mbox{\boldmath $L$}^\prime_\gamma &=& \int \,d^3 r \,
 E^k_\perp \,(\mbox{\boldmath $r$} \times \nabla) \,A^k_\perp .
\end{eqnarray}
The gauge-invariance of the first term can easily be convinced from the gauge
transformation property of the longitudinal component
\begin{eqnarray}
 \mbox{\boldmath $A$}_\parallel (\mbox{\boldmath $r$}_i) \ \rightarrow \ 
 \mbox{\boldmath $A$}_\parallel (\mbox{\boldmath $r$}_i) \ + \ 
 \nabla \Lambda (\mbox{\boldmath $r$}_i) ,
\end{eqnarray} 
combined with the gauge transformation property of quantum mechanical wave
function of charged particle system :
\begin{eqnarray}
 \Psi (\mbox{\boldmath $r$}_1, \cdots, \mbox{\boldmath $r$}_N) \ \rightarrow \ 
 \left(\, \prod_i^N \,\,e^{i \,q_i \,\Lambda (\mbox{\boldmath $r$}_i)} \,\right)
 \,\Psi (\mbox{\boldmath $r$}_1, \cdots \,\mbox{\boldmath $r$}_N) .
\end{eqnarray}
We emphasize that the pure-gauge covariant derivative in the Chen
formalism appears automatically. 
The gauge degrees of freedom, carried by the longitudinal component
is {\it not} introduced {\it by hand}.
It exists from the beginning in the original gauge theory !
This means that the Chen decomposition is not a GIE by the Stueckelberg trick.
Note however that the Chen decomposition is not only one GI decomposition.
Because the potential angular momentum $\mbox{\boldmath $J$}_{long}$
is solely gauge-invariant, we can leave it
in the photon OAM part, which leads to another GI decomposition, i.e. 
the decomposition (I), according to our classification \cite{ref:Waka11A}.

Another very important remark is as follows. It is a wide-spread belief that,
among the following two quantities, i.e. the canonical OAM and the mechanical OAM, 
\begin{eqnarray}
 \mbox{\boldmath $L$}_{can} = \mbox{\boldmath $r$} 
 \times \mbox{\boldmath $p$} \ \ \ \Leftrightarrow
 \ \ \ \mbox{\boldmath $L$}_{mech} = \mbox{\boldmath $r$} \times 
 (\mbox{\boldmath $p$} - e \,\mbox{\boldmath $A$}_\perp),
\end{eqnarray}
what is closer to physical image of orbital motion is the
former, because the latter appears to contain an {\it extra interaction term}
with the gauge field. This is a totally mistaken idea. In fact, the truth is just
opposite. We have shown above that the ``canonical'' OAM is a sum of the mechanical
OAM and the potential angular momentum as
\begin{eqnarray}
 \ \ \ \ \ \ \mbox{\boldmath $L$}_{``can"} &=&  
 \mbox{\boldmath $L$}_{mech}  
 \, + \, \sum_i \,
 \mbox{\boldmath $r$}_i \times q_i \,
 \mbox{\boldmath $A$}_\perp (\mbox{\boldmath $r$}_i)
 \, = \, \sum_i \,m_i \,\mbox{\boldmath $r$}_i \times \dot{\mbox{\boldmath $r$}}_i 
 \, + \, \int \,d^3 r \,\mbox{\boldmath $r$} \times 
 (\mbox{\boldmath $E$}_\parallel \times \mbox{\boldmath $B$}_\perp) .
\end{eqnarray}
As is clear from the expression of
mechanical OAM given as an outer product of $\mbox{\boldmath $r$}$ and $\dot{\mbox{\boldmath $r$}} = \mbox{\boldmath $v$}$, it is the
``mechanicalh OAM not the ``canonicalh
OAM that has a natural physical interpretation as orbital motion of particles.
It may sound paradoxical, but what contains an extra interaction term is rather
the ``canonicalhangular momentum than the ``mechanicalh angular momentum.    

As already pointed out, Lorc\'{e} claims that the decomposition of the gauge field
into the physical and pure-gauge components is not unique because
of the presence of the hidden St\"{u}ckelberg-like symmetry, which changes both
of $A^\mu_{phys}$ and $A^\mu_{pure}$ while keeping their sum
intact \cite{ref:Lorce12}.
This contradicts the above-explained common knowledge of electrodynamics that
the transverse-longitudinal decomposition is unique once the Lorentz-frame
of reference is specified. In the noncovariant treatment, the St\"{u}ckelberg
transformation introduced by Lorc\'{e} corresponds to a simultaneous transformation
of $\mbox{\boldmath $A$}_\parallel$ and $\mbox{\boldmath $A$}_\perp$ :
\begin{eqnarray}
 \mbox{\boldmath $A$}_\parallel \ \rightarrow \ \mbox{\boldmath $A$}^g_\parallel
 \ = \ \mbox{\boldmath $A$}_\parallel \ - \ \nabla \,C(x), \ \ \ \ \ \ 
 \mbox{\boldmath $A$}_\perp &\rightarrow& \mbox{\boldmath $A$}^g_\perp
 \ = \ \mbox{\boldmath $A$}_\perp \ + \ \nabla \,C(x), \label{eq:StueckelbergTr} 
\end{eqnarray}
with $C (x)$ being an arbitrary function of space-time.
(The similarity and the vital difference between the above St\"{u}ckelberg
transformation (\ref{eq:StueckelbergTr}) and the standard gauge transformation
(\ref{eq:GaugeTr}) should be clearly recognized.)
It was argued that, since this transformation alters both of
$\mbox{\boldmath $A$}_\parallel$ and $\mbox{\boldmath $A$}_\perp$ while
keeping the sum of them is intact, there are infinitely many
decompositions of $\mbox{\boldmath $A$}$
into $\mbox{\boldmath $A$}_\parallel$ and $\mbox{\boldmath $A$}_\perp$.
Here is an important oversight. The above transformation certainly keeps
the irrotational-free condition for $\mbox{\boldmath $A$}_\parallel$, since
\begin{equation}
 \nabla \times \mbox{\boldmath $A$}^g_\parallel \ = \ \nabla \times
 (\mbox{\boldmath $A$}_\parallel \ - \ \nabla \,C(x)) \ = \ 0. 
\end{equation}
However, it does not maintain the divergence-free (transversality) condition
for $\mbox{\boldmath $A$}_\perp$, since
\begin{equation}
 \nabla \cdot \mbox{\boldmath $A$}^g_\parallel \ = \ \nabla \cdot
 (\mbox{\boldmath $A$}_\parallel \ + \ \nabla \,C(x)) \ = \ \Delta \,C(x) 
 \ \neq \ 0 ,  
\end{equation}
unless $\Delta C(x) \ = \ 0$.
In the usual circumstances of electrodynamics, the harmonic function $C(x)$
satisfying $\Delta C(x) = 0$ can be set equal to zero without loss of
generality owing to the Helmholtz theorem. Thus, the Stueckelberg symmetry
does not exist, and the transverse-longitudinal
decomposition is unique.

\section{What is needed to settle the controversies}

We have shown that each term of our nucleon spin decomposition (I) and (II) is
separately gauge invariant, as long as the two parts of the decomposition
of $A^\mu$ satisfy the conditions (\ref{eq:cond1})-(\ref{eq:cond3}) under
general color SU(3) gauge transformation.
The fact that we did not give explicit formula for $A^\mu_{phys}$ and
$A^\mu_{pure}$ caused misunderstandings, however.
To resolve this misapprehension, we emphasize again the fact that the underlying
physics idea implicit in this decomposition is the transverse-longitudinal decomposition. 
From the physical viewpoint, the massless gauge field has only 2 transverse
degrees of freedom, and the other components are not independent
dynamical degrees of freedom. 
As was pointed out before, however, the transverse-longitudinal decomposition
can be made, only after specifying a particular Lorentz frame.
Fortunately, there exists a convenient method, with which we can make this decomposition in a seemingly covariant form which is convenient for
perturbative calculations of Feynman diagrams.
The key is a introduction of  a constant 4-vector $n^\mu$.
A typical example is Coulomb gauge-type projector in QED case, which
projects out the physical components of the photon field as extensively
discussed by Lavelle and McMullan \cite{ref:LavelleMcMullan93} : 
\begin{equation}
 A^\mu_{phys} (x) \ = \ P^{\mu \nu}_{phys} \,A_{\nu} (x), \label{eq:physical}
\end{equation}
where the projection operator is given by
\begin{eqnarray}
 P^{\mu \nu}_{phys} \ = \ g^{\mu \nu} \ + \ 
 \frac{\partial^\mu \,\partial^\nu - 
 \partial \cdot n \,(\partial^\mu \,n^\nu + \partial^\nu \,n^\mu) + 
 n^\mu \,n^\nu \,\Box}{(\partial \cdot n)^2 \ - \ \Box} .
\end{eqnarray}
with $n^\nu = (1,0,0,0)$ being a temporal vector. One can easily check that
this  projection operator satisfies the transversality condition
$k_\mu \,P^{\mu \nu}_{phys} = P^{\mu \nu}_{phys} \,k_\nu = 0$. 
More convenient for our purpose is a general axial-gauge type projector
given as follows.
\begin{equation}
 P^{\mu \nu}_{phys} \ = \ g^{\mu \nu} \ - \ 
 \frac{\partial^\mu \,n^\nu + \partial^\nu \,n^\mu}{\partial \cdot n} \ + \ 
 \frac{n^\mu \,n^\nu \,\Box}{(\partial \cdot n)^2} . \label{eq:axialproj}
\end{equation}
Here, $n^\mu$ is an arbitrary constant 4-vector, which can be either
of time-like, light-like, or space-like one.
Note that the above projection operator also satisfies the transversality
condition $k_\mu \,P^{\mu \nu}_{phys} = P^{\mu \mu}_{phys} \,k_\nu = 0$.
This ensures that $A^\mu_{phys}$ is gauge-invariant in the case of
abelian gauge theory.
In the case of nonabelian gauge theory, however, the physical
(or transverse) component of the gluon field given by 
(\ref{eq:physical}) and (\ref{eq:axialproj}) satisfies the desired
covariant gauge transformation property only at the lowest order
in the gauge coupling constant. Accordingly, the gluon spin operator
\begin{equation}
 M^{\mu \nu \lambda}_{G-spin} \ = \ 
 2 \,\mbox{Tr} \,\left[ F^{\mu \lambda} \,A^\nu_{phys} \ + \ 
 F^{\nu \lambda} \,A^\mu_{phys} \right],
\end{equation}
in which $A^\nu_{phys} (x)$ and $A^\mu_{phys} (x)$ are replaced by this
approximate form is regarded as a lowest order expression of
more rigorously defined gluon spin operator, so that it is expected to be
used in the calculation of the corresponding anomalous dimension
at the 1-loop level. (See (\cite{ref:Waka13}), for more detail.) 
On the basis of this expression of the gluon spin operator containing
arbitrary 4-vector $n^\mu$, which is thought to specify the Lorentz
frame in which the transversality condition is given and also the
quantization of the gauge field is carried out, we have calculated the
1-loop anomalous dimension matrix for the quark and gluon spin operators
in the nucleon, to find that it reproduces the
standardly-known answer irrespectively of the choice of $n^\mu$.
This is thought to give a further evidence to the gauge-independence
of gluon spin {\it in a traditional sense}.

\section{What is a problem of GIE approach ?}

Lorc\'{e} and Pasquini gave a useful relation between OAM and Wigner
distribution \cite{ref:LorcePasquini11}.
However, gauge-invariant definition of Wigner distribution generally depends
on the path of gauge link.
Hatta showed that the LC-like path choice gives ``canonical'' OAM \cite{ref:Hatta12}.
On the other hand, Ji, Xiong, and Yuan argued that the straight path connecting
the relevant two space-time points gives ``dynamical" OAM \cite{ref:JiXiongYuan12}. 
What plays a crucial role in these formulations is the so-called gauge-links.
The Wigner distributions defined through such gauge-links are gauge-invariant
by construction, but they are generally path-dependent.
The idea of gauge-link is of general nature and has a long history.
Once, DeWitt tried to formulate the quantum electrodynamics in a
gauge-invariant way, i.e. without introducing gauge-dependent
potential \cite{ref:DeWitt62}.
However, it was recognized soon that, although the framework
is manifestly gauge-invariant it {\it does} depend on the choice of
path defining the gauge-invariant potential \cite{ref:Belinfante62}
\nocite{ref:Mandelstam62}-\cite{ref:RohrlichStrocchi65}.
It was also demonstrated that path-dependence is eventually a reflection of
the gauge-dependence \cite{ref:Yang85}.
This dictates that, if a quantity in question is
seemingly gauge-invariant but path-dependent, it is not a gauge-invariant
quantity in a {\it true} or {\it traditional} sense, so that it may not
correspond to observables.
Undoubtedly, the GIE approach is equivalent to the standard treatment of gauge theory, only when its extension by means of gauge link is path-independent.
By the standard treatment of the gauge theory, we mean the following.
Start with a gauge-invariant quantity or expression.
Fix gauge in response to necessity of practical calculation. 
Answer should be independent of gauge choice.

\section{Summary and conclusion}
We have argued that there exist two and only two physically inequivalent
gauge-invariant decompositions of the nucleon spin, in sharp contrast to
the conflicting viewpoint that there are infinitely many decompositions
of the nucleon spin. 
These two decompositions, which we call (I) and (II), are characterized
by two different OAMs for quarks and gluons, i.e. the ``dynamical" OAM and
the generalized ``canonical" OAM. 
We have established the fact that the dynamical OAMs of quarks and gluons appearing
in the decomposition (I) can in principle be extracted model-independently from
combined analysis of  GPD  and polarized PDF measurements \cite{ref:Waka11A}.

On the other hand, the observability of the OAM appearing in the decomposition (II),
i.e. the generalized gcanonical" OAM is not clear yet.
This is because, although the relation between the ``canonical" OAM and a
Wigner distribution is suggested, its path-dependence or path-independence
should be clarified more throughly.
Moreover, once quantum loop effects are included, the very existence of TMDs
as well as Wigner distributions satisfying gauge-invariance and factorization
(or universality) at the same time is under investigation.
Is process-independent extraction of canonical OAM possible ?
One must say that this is still a challenging open question.

\acknowledgments
The author greatly acknowledge many stimulating discussions with C. Lorc\'{e},
although our opinions still differ in several essential points.
He would also like to thank the organizers of the 3rd Workshop on the
QCD structure of the Nucleon (QCD-N'12) held in Bilbao, Spain for
their kind hospitalities during the workshop.


\begin{thebibliography}{0}
\bibitem{ref:Waka10} \BY{Wakamatsu~M.}
  \IN{Phys. Rev. D}{81}{2010}{114010}.
\bibitem{ref:Waka11A} \BY{Wakamatsu~M.}
  \IN{Phys. Rev. D}{83}{2011}{014012}.
\bibitem{ref:Waka11B} \BY{Wakamatsu~M.}
  \IN{Phys. Rev. D}{84}{2011}{037501}.
\bibitem{ref:Waka12} \BY{Wakamatsu~M.}
  \IN{Phys. Rev. D}{85}{2012}{114039}.
\bibitem{ref:Chen08} \BY{Chen~X.S., L\"{u}~X.F., Sun~W.M., Wang~F. 
  \atque Goldman~T.}
  \IN{Phys. Rev. Lett.}{100}{2008}{232002}.
\bibitem{ref:Chen09} \BY{Chen~X.S., Sun~W.M., L\"{u}~X.F., Wang~F. 
  \atque Goldman~T.}
  \IN{Phys. Rev. Lett.}{103}{2009}{0620011}.
\bibitem{ref:Lorce12} \BY{Lorc\'{e}~C.}
  preprint arXiv:1205.6483 [hep-ph] ; arXiv:1210.2581 [hep-ph].
\bibitem{ref:Ji12} \BY{Ji~X, Xu~Y., \atque Zhao~Y.}
  \IN{JHEP}{1208}{2012}{082}.
\bibitem{ref:CohenT89} \BY{Cohen-Tanouji~C., Dupont-Roc~J. \atque Grynberg~G.}
  \TITLE{Photon \& Atoms},
       (Wiley, New York) 1989.
\bibitem{ref:LavelleMcMullan93} \BY{Lavelle~M., \atque McMullan~D.}
  \IN{Phys. Lett. B}{312}{1993}{211}.
\bibitem{ref:Waka13} \BY{Wakamatsu~M.}
  preprint arXiv:11302.5152 [hep-ph].
\bibitem{ref:LorcePasquini11} \BY{Lorc\'{e}~C., \atque Pasquini~B.}
  \IN{Phys. Rev. D}{84}{2011}{014015}.
\bibitem{ref:Hatta12} \BY{Hatta~Y.}
  \IN{Phys. Lett. B}{708}{2012}{186}.
\bibitem{ref:JiXiongYuan12} \BY{Ji~X, Xiong~X., \atque Yuan~F.}
  \IN{Phys. Rev. Lett.}{109}{2012}{152005}.
\bibitem{ref:DeWitt62} \BY{DeWitt~B.S.}
  \IN{Phys. Rev.}{125}{1962}{2189}.
\bibitem{ref:Belinfante62} \BY{Belinfante~F.}
  \IN{Phys. Rev. D}{128}{1962}{2832}.
\bibitem{ref:Mandelstam62} \BY{Mandelstam~S.}
  \IN{Ann. Phys.}{19}{1962}{1}.
\bibitem{ref:RohrlichStrocchi65} \BY{Rohrlich~F, \atque Strocchi~F.}
  \IN{Phys. Rev.}{139}{1965}{B476}.
\bibitem{ref:Yang85} \BY{Yang~K.-H.}
  \IN{J. Phys. A : Math Gen.}{18}{1985}{979}.


\end{thebibliography}
\end{document}